\documentclass[onecolumn,superscriptaddress,11pt,accepted=2018-09-14]{quantumarticle}
\pdfoutput=1
\usepackage[utf8]{inputenc}
\usepackage[english]{babel}
\usepackage[T1]{fontenc}
\usepackage{amsmath}
\usepackage{hyperref}
\usepackage{amsmath}
\usepackage{amssymb}
\usepackage{tikz}
\usepackage{lipsum}
\usepackage{mathtools}
\usepackage{dsfont}
\usepackage{amsthm}
\usepackage[numbers,sort&compress]{natbib}

\newtheorem{theorem}{Theorem}

\newtheorem{cor}[theorem]{Corollary}
\theoremstyle{definition}

\newtheorem*{remark}{Remark}

\newcommand{\ip}[2]{\langle #1 , #2\rangle}
\newcommand{\bigip}[2]{\bigl\langle #1, #2 \bigr\rangle}

\newcommand{\ket}[1]{\ensuremath{\lvert #1 \rangle}} %
\newcommand{\bra}[1]{\ensuremath{\langle #1 \rvert}} %

\newcommand\complex{\mathbb{C}}
\newcommand\real{\mathbb{R}}

\newcommand{\tinyspace}{\mspace{1mu}}

\newcommand{\op}[1]{\operatorname{#1}}
\newcommand{\tr}{\operatorname{Tr}}

\renewcommand{\vec}{\operatorname{vec}}
\newcommand{\fid}{\operatorname{F}}

\renewcommand{\t}{{\scriptscriptstyle\mathsf{T}}}

\newcommand{\norm}[1]{\lVert\tinyspace #1 \tinyspace\rVert}

\newcommand\I{\mathds{1}}

\newcommand{\setft}[1]{\mathrm{#1}}
\newcommand{\Density}{\setft{D}}
\newcommand{\Pos}{\setft{Pos}}

\newcommand{\Unitary}{\setft{U}}
\newcommand{\Herm}{\setft{Herm}}
\newcommand{\Lin}{\setft{L}}
\newcommand{\Channel}{\setft{C}}
\newcommand{\CP}{\setft{CP}}

\newcommand{\reg}[1]{\mathsf{#1}}

\newcommand\X{\mathcal{X}}
\newcommand\Y{\mathcal{Y}}
\newcommand\Z{\mathcal{Z}}
\newcommand\W{\mathcal{W}}

\renewcommand\S{\mathcal{S}}

\definecolor{White}{rgb}{1,1,1}
\definecolor{Black}{rgb}{0,0,0}
\definecolor{LightGray}{rgb}{.81,.81,.81}
\colorlet{ChannelColor}{LightGray}
\colorlet{ChannelTextColor}{Black}
\colorlet{ReadoutColor}{White}

\begin{document}

\title{Time-reversal of rank-one quantum strategy\newline functions}
\date{October 1, 2018}

\author{Yuan Su}
\affiliation{
  Department of Computer Science,
  Institute for Advanced Computer Studies, and
  Joint Center for Quantum Information and Computer Science,
  University of Maryland, USA\vspace{1mm}}

\homepage{http://quics.umd.edu/people/yuan-su}
\orcid{0000-0003-1144-3563}

\author{John Watrous}
\affiliation{Institute for Quantum Computing and School of Computer
  Science, University of Waterloo, Canada\vspace{1mm}}

\affiliation{Canadian Institute for Advanced Research, Toronto,
  Canada\vspace{2mm}}

\makeatletter
\@homepage{}{%
  \href{https://cs.uwaterloo.ca/~watrous/}{%
    https://cs.uwaterloo.ca/{\raise.17ex\hbox{$\scriptstyle\sim$}}watrous/}}
\makeatother

\orcid{0000-0002-4263-9393}

\maketitle

\begin{abstract}
  The \emph{quantum strategy} (or \emph{quantum combs}) framework is a useful
  tool for reasoning about interactions among entities that process and
  exchange quantum information over the course of multiple turns.
  We prove a time-reversal property for a class of linear functions, defined
  on quantum strategy representations within this framework, that corresponds
  to the set of rank-one positive semidefinite operators on a certain space.
  This time-reversal property states that the maximum value obtained by such a
  function over all valid quantum strategies is also obtained when the
  direction of time for the function is reversed, despite the fact that the
  strategies themselves are generally not time reversible.
  An application of this fact is an alternative proof of a known relationship
  between the conditional min- and max-entropy of bipartite quantum states,
  along with generalizations of this relationship.
\end{abstract}


\section{The quantum strategy framework}

The \emph{quantum strategy framework} \cite{GutoskiW07}, which is also known as
the \emph{quantum combs framework} \cite{ChiribellaDP08,ChiribellaDP09},
provides a useful framework for reasoning about networks of quantum channels.
It may be used to model scenarios in which two or more entities, which we will
call \emph{players}, process and exchange quantum information over the course
of multiple rounds of communication; and it is particularly useful when one
wishes to consider an optimization over all possible behaviors of one
player, for any given specification of the other player or players.
Various developments, applications, and variants of the quantum strategy
framework can be found in
\cite{ChiribellaDP08b,ChiribellaDPSW13,ChiribellaE16,Gutoski09,Hardy12},
for instance, and in a number of other sources.

In the discussion of the quantum strategy framework that follows, as well as in
the subsequent sections of this paper, we assume that the reader is familiar
with quantum information theory and semidefinite programming.
References on this material include
\cite{NielsenC00, Wilde13, KitaevSV02, WolkowiczSV00} as well as
\cite{Watrous18}, which we follow closely with respect to notation and
terminology.
In particular, we denote quantum registers by capital sans serif letters such
as $\reg{X}$, $\reg{Y}$, and $\reg{Z}$ (sometimes with natural number
subscripts), while the same letters (with matching subscripts) in a scripted
font, such as $\X$, $\Y$, and $\Z$ denote the complex Euclidean spaces
(i.e., finite-dimensional complex Hilbert spaces) associated with the
corresponding registers.
The set $\Lin(\X,\Y)$ denotes the set of all linear operators from $\X$ to
$\Y$;
$\Lin(\X)$ is a shorthand for $\Lin(\X,\X)$;
$\Herm(\X)$, $\Pos(\X)$, $\Density(\X)$, and $\Unitary(\X)$ denote the sets
of all Hermitian operators, positive semidefinite operators, density operators,
and unitary operators acting on $\X$;
$\Channel(\X,\Y)$ denotes the set of all channels (i.e., completely positive and
trace-preserving maps) mapping $\Lin(\X)$ to $\Lin(\Y)$; and
$\Channel(\X)$ is a shorthand for $\Channel(\X,\X)$.
The adjoint of an operator $A$ is denoted $A^{\ast}$, the entry-wise complex
conjugate is denoted $\overline{A}$, and the transpose is denoted $A^{\t}$.
A similar notation is used for the adjoint and transpose of a channel $\Phi$
(the meaning of which, in the case of the transpose, will be clarified later).
The (Hilbert-Schmidt) inner-product is defined as $\ip{A}{B} = \tr(A^{\ast}B)$
for all operators $A,B\in\Lin(\X)$.
Some additional notation will be introduced as it is used.

\subsection*{An example of a six-message interaction}

To explain the aspects of the quantum strategy framework that are relevant to
this paper, we will begin by discussing an example of an interaction structure
involving six messages exchanged between two players, Alice and Bob.
We have chosen to describe a six-message interaction because it is simple and
concrete, but nevertheless clearly suggests the underlying structure of an
interaction having any finite number of message exchanges.
Our main result holds in the general case, which will be considered later,
where an arbitrary finite number of message exchanges may take place.

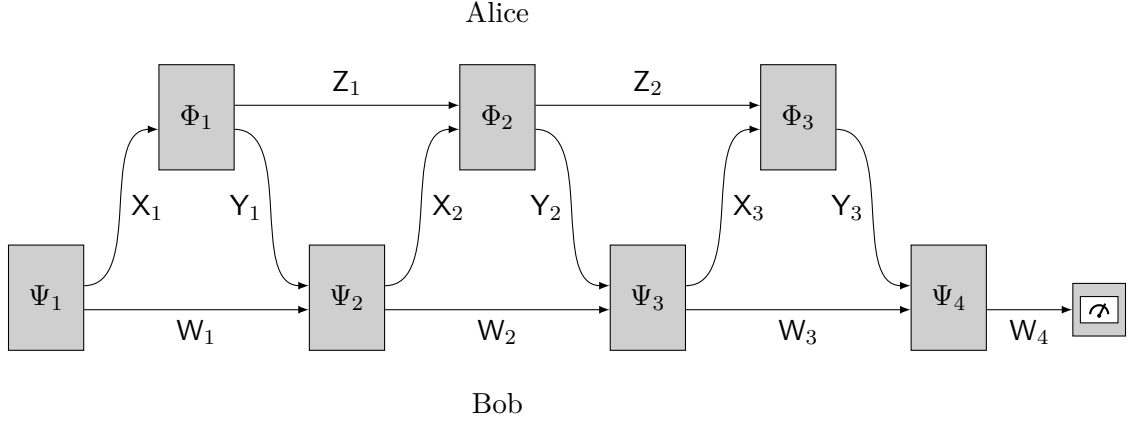
\begin{figure}[t]
  \begin{center}
    \begin{tikzpicture}[scale=0.4, 
        turn/.style={draw, minimum height=14mm, minimum width=10mm,
          fill = ChannelColor, text=ChannelTextColor},
        measure/.style={draw, minimum width=7mm, minimum height=7mm,
          fill = ChannelColor},
        invisible/.style={minimum height=1mm, minimum width=10mm},
        >=latex]
      
      \node (B1) at (-18,-4) [turn] {$\Psi_1$};
      \node (B2) at (-8,-4) [turn] {$\Psi_2$};
      \node (B3) at (2,-4) [turn] {$\Psi_3$};
      \node (B4) at (12,-4) [turn] {$\Psi_4$};
      
      \node (M) at (17,-4.4) [measure] {};
     
      \node (A1) at (-13,2) [turn] {$\Phi_1$};
      \node (A2) at (-3,2) [turn] {$\Phi_2$};
      \node (A3) at (7,2) [turn] {$\Phi_3$};

      \node[draw, minimum width=5mm, minimum height=3.5mm, fill=ReadoutColor]
      (readoutA) at (M) {};

      \draw[thick] ($(M)+(0.3,-0.15)$) arc (0:180:3mm);
      \draw[thick] ($(M)+(0.2,0.2)$) -- ($(M)+(0,-0.2)$);
      \draw[fill] ($(M)+(0,-0.2)$) circle (0.5mm);
      
      \draw[->] ([yshift=-4mm]B1.east) -- ([yshift=-4mm]B2.west)
      node [below, midway] {$\reg{W}_1$};
      
      \draw[->] ([yshift=-4mm]B2.east) -- ([yshift=-4mm]B3.west)
      node [below, midway] {$\reg{W}_2$};
      
      \draw[->] ([yshift=-4mm]B3.east) -- ([yshift=-4mm]B4.west)
      node [below, midway] {$\reg{W}_3$};

      \draw[->] ([yshift=-4mm]B4.east) -- (M.west)
      node [below, midway] {$\reg{W}_4$};
      
      \draw[->] ([yshift=4mm]B1.east) .. controls +(right:20mm) and
      +(left:20mm) .. ([yshift=-4mm]A1.west) node [right, pos=0.5]
      {$\reg{X}_1$};
      
      \draw[->] ([yshift=4mm]B2.east) .. controls +(right:20mm) and 
      +(left:20mm) .. ([yshift=-4mm]A2.west) node [right, pos=0.5]
      {$\reg{X}_2$};
      
      \draw[->] ([yshift=4mm]B3.east) .. controls +(right:20mm) and 
      +(left:20mm) .. ([yshift=-4mm]A3.west) node [right, pos=0.5]
      {$\reg{X}_3$};
      
      \draw[->] ([yshift=-4mm]A1.east) .. controls +(right:20mm) and 
      +(left:20mm) .. ([yshift=4mm]B2.west) node [left, pos=0.5] {$\reg{Y}_1$};
      
      \draw[->] ([yshift=-4mm]A2.east) .. controls +(right:20mm) and 
      +(left:20mm) .. ([yshift=4mm]B3.west) node [left, pos=0.5] {$\reg{Y}_2$};
      
      \draw[->] ([yshift=-4mm]A3.east) .. controls +(right:20mm) and 
      +(left:20mm) .. ([yshift=4mm]B4.west) node [left, pos=0.5] {$\reg{Y}_3$};
      
      \draw[->] ([yshift=4mm]A1.east) -- ([yshift=4mm]A2.west)
      node [above, midway] {$\reg{Z}_1$};
      
      \draw[->] ([yshift=4mm]A2.east) -- ([yshift=4mm]A3.west)
      node [above, midway] {$\reg{Z}_2$};
      
      \node at (-3,5.5) {Alice};
      \node at (-3,-7.5) {Bob};

    \end{tikzpicture}
  \end{center}
  \caption{A six message interaction between Alice and Bob, after which Bob
    produces a measurement outcome.}
  \label{fig:Alice-and-Bob-interact}
\end{figure}

Figure~\ref{fig:Alice-and-Bob-interact} illustrates an interaction between
Alice and Bob.
In this figure, time proceeds from left to right, and the arrows represent
registers either being sent from one player to the other
(as is the case for the registers $\reg{X}_1$, $\reg{Y}_1$, $\reg{X}_2$,
$\reg{Y}_2$, $\reg{X}_3$, and $\reg{Y}_3$), or momentarily stored by one of the
two players (as is the case for $\reg{Z}_1$ and $\reg{Z}_2$, stored by Alice,
and $\reg{W}_1$, $\reg{W}_2$, $\reg{W}_3$, $\reg{W}_4$, stored by Bob).
Alice's actions are represented by the channels $\Phi_1$, $\Phi_2$, and
$\Phi_3$, and Bob's actions are represented by the channels $\Psi_1$, $\Psi_2$,
$\Psi_3$, and $\Psi_4$, as well as a final measurement, which is not given a
name in the figure.

Suppose that Bob's specification has been fixed, including his choices for the
channels $\Psi_1$, $\Psi_2$, $\Psi_3$, and $\Psi_4$, as well as his final
measurement, and suppose further that one of Bob's possible measurement
outcomes is to be viewed as desirable to Alice.
It is then natural to consider an optimization over Alice's possible actions,
maximizing the probability that Bob's measurement produces the outcome Alice
desires.
The quantum strategy framework reveals that this optimization problem can be
expressed as a semidefinite program, in the manner that will now be described.

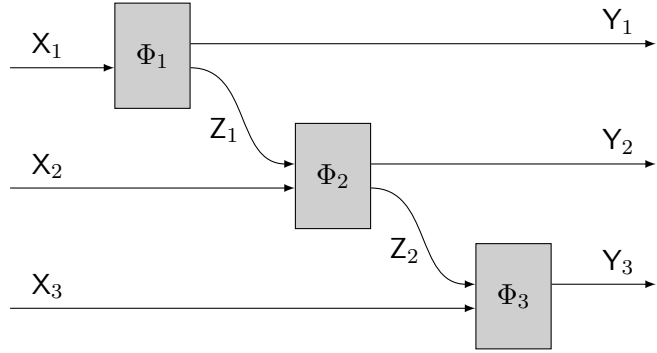
\begin{figure}[t]
  \begin{center}
    \begin{tikzpicture}[scale=0.4, 
        turn/.style={draw, minimum height=14mm, minimum width=10mm,
          fill = ChannelColor, text=ChannelTextColor},
        measure/.style={draw, minimum width=7mm, minimum height=7mm,
          fill = ChannelColor},
        invisible/.style={minimum height=1mm, minimum width=10mm},
        >=latex]
      
      \node (X1) at (-12,4) [invisible] {};
      \node (X2) at (-12,0) [invisible] {};
      \node (X3) at (-12,-4) [invisible] {};
      
      \node (Y1) at (12,4) [invisible] {};
      \node (Y2) at (12,0) [invisible] {};
      \node (Y3) at (12,-4) [invisible] {};
     
      \node (B1) at (-6,4) [turn] {$\Phi_1$};
      \node (B2) at (0,0) [turn] {$\Phi_2$};
      \node (B3) at (6,-4) [turn] {$\Phi_3$};

      \draw[->] ([yshift=-4mm]X1.east) -- ([yshift=-4mm]B1.west)
      node [above, xshift=5mm, pos=0] {$\reg{X}_1$};

      \draw[->] ([yshift=-4mm]X2.east) -- ([yshift=-4mm]B2.west)
      node [above, xshift=5mm, pos=0] {$\reg{X}_2$};

      \draw[->] ([yshift=-4mm]X3.east) -- ([yshift=-4mm]B3.west)
      node [above, xshift=5mm, pos=0] {$\reg{X}_3$};

      \draw[->] ([yshift=4mm]B1.east) -- ([yshift=4mm]Y1.west)
      node [above, xshift=-5mm, pos=1] {$\reg{Y}_1$};

      \draw[->] ([yshift=4mm]B2.east) -- ([yshift=4mm]Y2.west)
      node [above, xshift=-5mm, pos=1] {$\reg{Y}_2$};

      \draw[->] ([yshift=4mm]B3.east) -- ([yshift=4mm]Y3.west)
      node [above, xshift=-5mm, pos=1] {$\reg{Y}_3$};
      
      \draw[->] ([yshift=-4mm]B1.east) .. controls +(right:20mm) and
      +(left:20mm) .. ([yshift=4mm]B2.west) node [left, pos=0.6]
      {$\reg{Z}_1$};
      
      \draw[->] ([yshift=-4mm]B2.east) .. controls +(right:20mm) and 
      +(left:20mm) .. ([yshift=4mm]B3.west) node [left, pos=0.6]
      {$\reg{Z}_2$};

    \end{tikzpicture}
  \end{center}
  \caption{The channel $\Xi_3$ that describes Alice's actions in the
    interaction illustrated in Figure~\ref{fig:Alice-and-Bob-interact}.}
  \label{fig:Alice-as-channel}
\end{figure}

First, a single channel $\Xi_3$ that transforms
$(\reg{X}_1,\reg{X}_2,\reg{X}_3)$ to $(\reg{Y}_1,\reg{Y}_2,\reg{Y}_3)$ is
associated with any given choice for Alice's actions.
That is, the channel $\Xi_3$ takes the form
\begin{equation}
  \label{eq:3-channel-form}
  \Xi_3\in\Channel(\X_1\otimes\X_2\otimes\X_3,\Y_1\otimes\Y_2\otimes\Y_3),
\end{equation}
and for a particular selection of $\Phi_1$, $\Phi_2$, and $\Phi_3$ may be
expressed as
\begin{equation}
  \label{eq:3-channel-composition}
  \Xi_3 = \bigl(\I_{\Lin(\Y_1\otimes\Y_2)} \otimes \Phi_3\bigr)
  \bigl(\I_{\Lin(\Y_1)} \otimes \Phi_2 \otimes \I_{\Lin(\X_3)}\bigr)
  \bigl(\Phi_1 \otimes \I_{\Lin(\X_2 \otimes \X_3)}\bigr).
\end{equation}
Formally speaking, this composition requires that we view $\Phi_1$, $\Phi_2$,
and $\Phi_3$ as channels of the form
$\Phi_1\in\Channel(\X_1,\Y_1\otimes\Z_1)$,
$\Phi_2 \in \Channel(\Z_1\otimes\X_2,\Y_2\otimes\Z_2)$, and
$\Phi_3 \in \Channel(\Z_2\otimes\X_3,\Y_3)$,
as opposed to the forms
$\Phi_1\in\Channel(\X_1,\Z_1\otimes\Y_1)$,
$\Phi_2 \in \Channel(\Z_1\otimes\X_2,\Z_2\otimes\Y_2)$, and
$\Phi_3 \in \Channel(\Z_2\otimes\X_3,\Y_3)$ suggested by
Figure~\ref{fig:Alice-and-Bob-interact}, so that the ordering of the tensor
factors of the various input and output spaces is consistent with the
composition.
Similar re-orderings of tensor factors should be assumed implicitly throughout
this paper as needed.
This understanding should not be a source of confusion because we always
assign distinct names to distinct registers (and their associated spaces).
Figure~\ref{fig:Alice-as-channel} illustrates the action of the channel
$\Xi_3$, which in words may be described as the channel obtained if all three
of the registers $(\reg{X}_1,\reg{X}_2,\reg{X}_3)$ are provided initially, and
then Alice's actions are composed in the natural way to produce
$(\reg{Y}_1,\reg{Y}_2,\reg{Y}_3)$ as output registers.

It may appear that by considering the channel $\Xi_3$, one is ignoring the
possibility that Bob's actions could, for instance, allow the contents of
$\reg{Y}_1$ or $\reg{Y}_2$ to influence what is input into $\reg{X}_2$ or
$\reg{X}_3$.
Despite this appearance, the influence that Alice's actions have from the
viewpoint of Bob, including the probability for each of his measurement
outcomes to appear, is uniquely determined by the channel $\Xi_3$.

Naturally, not all channels of the form \eqref{eq:3-channel-form} will arise
from a composition of channels $\Phi_1$, $\Phi_2$, and $\Phi_3$ as in
\eqref{eq:3-channel-composition}; the fact that $\Phi_1$ is effectively
performed first, $\Phi_2$ is performed second, and $\Phi_3$ is performed third
imposes constraints on the channels $\Xi_3$ that can be obtained.
In particular, consider the channel that results when $\Xi_3$ is performed
and then the partial trace is performed on $\Y_3$.
As $\Phi_3$ is a channel, discarding its output is equivalent to discarding its
inputs, from which it follows that
\begin{equation}
  \label{eq:channel-constraint-3}
  \text{Tr}_{\Y_3} \circ \Xi_3 = \Xi_2 \circ \text{Tr}_{\X_3},
\end{equation}
where the circles represent channel compositions and
$\Xi_2\in\Channel(\X_1\otimes\X_2,\Y_1\otimes\Y_2)$ is the channel defined as
\begin{equation}
  \Xi_2 = \bigl(\I_{\Lin(\Y_1)} \otimes (\textup{Tr}_{\Z_2} \circ \Phi_2)\bigr)
  \bigl(\Phi_1 \otimes \I_{\Lin(\X_2)}\bigr).
\end{equation}
That is, $\Xi_2$ is the channel obtained from $\Phi_1$ and $\Phi_2$, followed
by the partial trace over $\Z_2$, by a similar process to the one used to
obtain $\Xi_3$.
By similar reasoning, one finds that
\begin{equation}
  \label{eq:channel-constraint-2}
  \text{Tr}_{\Y_2} \circ \Xi_2 = \Xi_1 \circ \text{Tr}_{\X_2},
\end{equation}
where $\Xi_1\in\Channel(\X_1,\Y_1)$ is the channel given by
$\Xi_1 = \textup{Tr}_{\Z_1}\circ \Phi_1$.

Somewhat remarkably, this is not only a necessary condition on the channel
$\Xi_3$, but also a sufficient one, for it to be obtained from a composition of
channels $\Phi_1$, $\Phi_2$, and $\Phi_3$ as described above.
That is, given any channel
\begin{equation}
  \Xi_3\in\Channel(\X_1\otimes\X_2\otimes\X_3,\Y_1\otimes\Y_2\otimes\Y_3)
\end{equation}
satisfying \eqref{eq:channel-constraint-3} and \eqref{eq:channel-constraint-2},
for some choice of channels
\begin{equation}
  \begin{gathered}
    \Xi_2\in\Channel(\X_1\otimes\X_2,\Y_1\otimes\Y_2),\\
    \Xi_1\in\Channel(\X_1,\Y_1),
  \end{gathered}
\end{equation}
there must exist channels
\begin{equation}
  \begin{gathered}
    \Phi_1\in\Channel(\X_1,\Y_1\otimes\Z_1),\\
    \Phi_2\in\Channel(\Z_1\otimes\X_2,\Y_2\otimes\Z_2),\\
    \Phi_3\in\Channel(\Z_2\otimes\X_3,\Y_3),
  \end{gathered}
\end{equation}
for spaces $\Z_1$ and $\Z_2$ having sufficiently large dimension, so that
\eqref{eq:3-channel-composition} holds.
This fact is proved in \cite{ChiribellaDP08,ChiribellaDP09,GutoskiW07}, and we
note that a key idea through which this equivalence is proved may be found in
\cite{EggelingSW02}.

The next step toward an expression of the optimization problem suggested above
as a semidefinite program makes use of the Choi representation of channels.
The Choi representation of the channel $\Xi_3$ takes the form
\begin{equation}
  J(\Xi_3) \in
  \Pos(\Y_1\otimes\Y_2\otimes\Y_3\otimes\X_1\otimes\X_2\otimes\X_3),
\end{equation}
as the complete positivity of $\Phi_1$, $\Phi_2$, and $\Phi_3$ implies that
$\Xi_3$ is also completely positive, and therefore $J(\Xi_3)$ is positive
semidefinite.
The constraints on the channel $\Xi_3$ described previously correspond
(conveniently) to linear constraints;
one has that \eqref{eq:channel-constraint-3} and
\eqref{eq:channel-constraint-2} hold, for some choice of channels $\Xi_2$ and
$\Xi_1$, if and only if the Choi representation $X_3 = J(\Xi_3)$ of $\Xi_3$
satisfies
\begin{equation}
  \begin{gathered}
    \tr_{\Y_3}(X_3) = X_2 \otimes \I_{\X_3},\\
    \tr_{\Y_2}(X_2) = X_1 \otimes \I_{\X_2},\\
    \tr_{\Y_1}(X_1) = \I_{\X_1},
  \end{gathered}
\end{equation}
for some choice of operators
\begin{equation}
  \begin{gathered}
    X_2 \in \Pos(\Y_1\otimes\Y_2\otimes\X_1\otimes\X_2),\\
    X_1 \in \Pos(\Y_1\otimes\X_1).
  \end{gathered}
\end{equation}
These operators correspond to the Choi representations $X_2 = J(\Xi_2)$ and
$X_1 = J(\Xi_1)$.

Finally, the probability that Bob's measurement produces a given outcome
is a linear function of the channel $\Xi_3$, and is therefore a linear function
of the Choi representation $X_3 = J(\Xi_3)$.
Although this process is not relevant to the main result of this paper, we note
that it is possible to obtain an explicit description of this linear function
given a specification of Bob's actions, including his final measurement.
In somewhat vague terms, the linear function describing Bob's probability to
produce a particular measurement outcome is given by $\ip{P}{X_3}$, where
\begin{equation}
  P\in\Pos(\Y_1\otimes\Y_2\otimes\Y_3\otimes\X_1\otimes\X_2\otimes\X_3)
\end{equation}
is an operator that is obtained from $\Psi_1$, $\Psi_2$, $\Psi_3$, $\Psi_4$,
and the measurement operator corresponding to the outcome being considered
by a process very similar to the one through which $X_3$ is obtained from
$\Phi_1$, $\Phi_2$, and $\Phi_3$.
The reader is again referred to
\cite{GutoskiW07,ChiribellaDP08,ChiribellaDP09} for further details.

More generally, an arbitrary real-valued linear function of the operator
$X_3$ may be expressed as $\ip{H}{X_3}$ for some choice of a Hermitian operator
\begin{equation}
  H\in\Herm(\Y_1\otimes\Y_2\otimes\Y_3\otimes\X_1\otimes\X_2\otimes\X_3),
\end{equation}
which need not represent the probability with which a particular measurement
outcome is obtained for channels $\Psi_1,\ldots,\Psi_4$ followed by a
measurement.
Such a function could, for instance, represent an expected payoff for Alice's
actions, under the assumption that a real-valued payoff is associated with each
of Bob's measurement outcomes.

\subsection*{General semidefinite programming formulation}

As mentioned previously, the six-message example just described generalizes to
any finite number of message exchanges.
If the number of message exchanges is equal to $n$, the input registers to
Alice (the player whose actions are being optimized) are
$\reg{X}_1,\ldots,\reg{X}_n$, and the output registers of Alice are
$\reg{Y}_1,\ldots,\reg{Y}_n$, then the possible strategies for Alice are
represented by channels of the form
\begin{equation}
  \Xi_n \in \Channel(\X_1\otimes\cdots\otimes\X_n,
  \Y_1\otimes\cdots\otimes\Y_n)
\end{equation}
that obey constraints that generalize \eqref{eq:channel-constraint-3} and
\eqref{eq:channel-constraint-2}.
Specifically, there must exist channels
\begin{equation}
  \begin{gathered}
    \Xi_{n-1} \in \Channel(\X_1\otimes\cdots\otimes\X_{n-1},
    \Y_1\otimes\cdots\otimes\Y_{n-1})\\
    \vdots\\
    \Xi_1\in\Channel(\X_1,\Y_1)
  \end{gathered}
\end{equation}
such that
\begin{equation}
  \text{Tr}_{\Y_k} \circ \Xi_k = \Xi_{k-1} \circ \text{Tr}_{\X_k}
\end{equation}
for all $k\in\{2,\ldots,n\}$.
For the maximization of a real-valued linear function over all strategies for
Alice, represented by a Hermitian operator
\begin{equation}
  H\in\Herm(\Y_1\otimes\cdots\otimes\Y_n\otimes\X_1\otimes\cdots\otimes\X_n),
\end{equation}
one obtains the semidefinite program described in
Figure~\ref{figure:semidefinite-program}.
The primal problem corresponds to an optimization over all Choi
representations of the channels $\Xi_1,\ldots,\Xi_n$.
This semidefinite programming formulation is implicit in \cite{GutoskiW07},
and first appeared explicitly in \cite{Gutoski09}.
It also appears in \cite{ChiribellaE16}, where it was used to define
a generalized notion of min-entropy for quantum networks.

\begin{figure}[!t]
  \begin{center}
    \centerline{\underline{Primal problem}}\vspace{-6mm}
    \begin{align*}
      \text{maximize:}\quad & \ip{H}{X_n}\\
      \text{subject to:}\quad 
      & \tr_{\Y_n}(X_n) = X_{n-1} \otimes \I_{\X_n},\\
      & \qquad\vdots\\
      & \tr_{\Y_2}(X_2) = X_1 \otimes \I_{\X_2},\\
      & \tr_{\Y_1}(X_1) = \I_{\X_1},\\
      & X_n \in \Pos(\Y_1\otimes\cdots\otimes\Y_n\otimes\X_1\otimes\cdots
      \otimes\X_n),\\
      & \qquad \vdots\\
      & X_2 \in \Pos(\Y_1\otimes\Y_2\otimes\X_1\otimes\X_2),\\
      & X_1 \in \Pos(\Y_1\otimes\X_1).
    \end{align*}\\[2mm]
    \centerline{\underline{Dual problem}}\vspace{-6mm}
    \begin{align*}
      \text{minimize:}\quad & \tr(Y_1)\\
      \text{subject to:}\quad 
      & Y_n \otimes \I_{\Y_n} \geq H,\\
      & Y_{n-1} \otimes \I_{\Y_{n-1}} \geq \tr_{\X_n}(Y_n),\\
      & \qquad\vdots\\
      & Y_1 \otimes \I_{\Y_1} \geq \tr_{\X_2}(Y_2),\\
      & Y_n \in \Herm(\Y_1\otimes\cdots\otimes\Y_{n-1}\otimes\X_1\otimes\cdots
      \otimes\X_n),\\
      & Y_{n-1} \in \Herm(\Y_1\otimes\cdots\otimes\Y_{n-2}\otimes
      \X_1\otimes\cdots\otimes\X_{n-1}),\\
      & \qquad \vdots\\
      & Y_1 \in \Herm(\X_1).
    \end{align*}
  \end{center}
  \caption{The semidefinite program representing a maximization of a linear
    function of an $n$-turn strategy.}
  \label{figure:semidefinite-program}
\end{figure}

It may be noted that the general problem just formulated concerns interactions
involving an even number of register exchanges, where Alice (the player whose
actions are being optimized) always receives the first transmission,
represented by $\reg{X}_1$, and sends the last transmission, represented by
$\reg{Y}_n$.
However, one is free to take either or both of the registers $\reg{X}_1$ and
$\reg{Y}_n$ to be trivial registers, so that correspondingly $\X_1 = \complex$
and/or $\Y_n = \complex$.
This is tantamount to allowing either an odd number of register exchanges or an
even number in the situation that Alice sends the first (nontrivial) register
and receives the last.

\section{Statement and proof of the main result}

The main result of the current paper concerns the optimization problem
described in the previous section, as represented by the semidefinite program
in Figure~\ref{figure:semidefinite-program}, in the case that $H = u u^{\ast}$
is a rank one positive semidefinite operator.
The result to be described does not hold in general when $H$ does not take this
form.

In order to explain the main result in precise terms, it will be helpful to
introduce some notation.
Suppose that a positive integer $n$ along with spaces $\X_1,\ldots,\X_n$ and
$\Y_1,\ldots,\Y_n$ have been fixed.
For each $k\in\{1,\ldots,n\}$, let
\begin{equation}
  \S_k(\X_1,\ldots,\X_k; \Y_1,\ldots,\Y_k)
  \subset \Pos(\Y_1\otimes\cdots\otimes\Y_k\otimes\X_1\otimes\cdots\otimes\X_k)
\end{equation}
denote the primal-feasible choices for the operator $X_k$ in the semidefinite
program specified in Figure~\ref{figure:semidefinite-program}.
That is, we define
\begin{equation}
  \S_1(\X_1;\Y_1) = \bigl\{X_1\in\Pos(\Y_1\otimes\X_1)\,:\,
  \tr_{\Y_1}(X_1) = \I_{\X_1}\bigr\}
\end{equation}
(which is the set of all Choi operators of channels of the form
$\Xi_1 \in \Channel(\X_1,\Y_1)$), and
\begin{equation}
  \begin{multlined}
    \S_k(\X_1,\ldots,\X_k;\Y_1,\ldots,\Y_k)\\[2mm]
    = \bigl\{
    X_k\in\Pos(\Y_1\otimes\cdots\otimes\Y_k\otimes\X_1\otimes\cdots\otimes\X_k)
    \,:\,\tr_{\Y_k}(X_k) = X_{k-1}\otimes\I_{\X_k}\\
    \text{for some}\;X_{k-1}\in\S_{k-1}(\X_1,\ldots,
    \X_{k-1};\Y_1,\ldots,\Y_{k-1})\bigr\}
  \end{multlined}
\end{equation}
for $k \in \{2,\ldots,n\}$.
The primal form of the semidefinite program described in
Figure~\ref{figure:semidefinite-program} can therefore be expressed succinctly
as
\begin{equation}
  \begin{aligned}
    \text{maximize:} \quad & \ip{H}{X}\\
    \text{subject to:} \quad & X\in
    \S_n(\X_1,\ldots,\X_n;\Y_1,\ldots,\Y_n).
  \end{aligned}
\end{equation}
We will refer to operators in the sets defined above as
\emph{strategy operators}, as they represent $n$-turn strategies with respect
to the quantum strategy framework.

Let us also define an isometry
\begin{equation}
  W \in
  \Unitary(\Y_1\otimes\cdots\otimes\Y_n\otimes\X_1\otimes\cdots\otimes\X_n,
  \X_n\otimes\cdots\otimes\X_1\otimes\Y_n\otimes\cdots\otimes\Y_1)
\end{equation}
by the action
\begin{equation}
  W (y_1\otimes\cdots\otimes y_n\otimes x_1\otimes\cdots\otimes x_n)
  = x_n\otimes\cdots\otimes x_1\otimes y_n\otimes\cdots\otimes y_1
\end{equation}
for all vectors $x_1\in\X_1,\ldots,x_n\in\X_n$ and
$y_1\in\Y_1,\ldots,y_n\in\Y_n$.
In words, $W$ simply reverses the order of the tensor factors of the space
$\Y_1\otimes\cdots\otimes\Y_n\otimes\X_1\otimes\cdots\otimes\X_n$, yielding a
vector in $\X_n\otimes\cdots\otimes\X_1\otimes\Y_n\otimes\cdots\otimes\Y_1$
that, aside from this re-ordering of tensor factors, is the same as its input
vector.

\subsection*{Statement of the main result}

With the notation just introduced in hand, the main theorem may now be stated.

\begin{theorem}
  \label{theorem:main}
  Let $\X_1,\ldots,\X_n$ and $\Y_1,\ldots,\Y_n$ be complex Euclidean spaces,
  for $n$ a positive integer, let
  \begin{equation}
    u \in \Y_1\otimes\cdots\otimes\Y_n\otimes\X_1\otimes\cdots\otimes\X_n
  \end{equation}
  be a vector, and let
  \begin{equation}
    X\in\S_n(\X_1,\ldots,\X_n;\Y_1,\ldots,\Y_n)
  \end{equation}
  be a strategy operator.
  There exists a strategy operator
  \begin{equation}
    Y\in\S_n(\Y_n,\ldots,\Y_1;\X_n,\ldots,\X_1)
  \end{equation}
  such that
  \begin{equation}
    \label{eq:main-theorem-inequality}
    \ip{W u u^{\ast} W^{\ast}}{Y} \geq \ip{u u^{\ast}}{X}.
  \end{equation}
  If it is the case that
  $\dim(\Y_1\otimes\cdots\otimes\Y_n)\leq\dim(\X_1\otimes\cdots\otimes\X_n)$,
  then the operator $Y$ may be chosen so that equality holds in
  \eqref{eq:main-theorem-inequality}.
\end{theorem}

\begin{cor}
  \label{cor:main}
  Let $\X_1,\ldots,\X_n$ and $\Y_1,\ldots,\Y_n$ be complex Euclidean spaces,
  for $n$ a positive integer, and let
  \begin{equation}
    u \in \Y_1\otimes\cdots\otimes\Y_n\otimes\X_1\otimes\cdots\otimes\X_n
  \end{equation}
  be a vector.
  The semidefinite optimization problems
  \begin{equation}
    \begin{aligned}
      \text{maximize:}\quad & \ip{u u^{\ast}}{X}\\
      \text{subject to:}\quad 
      & X \in \S_n(\X_1,\ldots,\X_n;\Y_1,\ldots,\Y_n)
    \end{aligned}
  \end{equation}
  and
  \begin{equation}
    \begin{aligned}
      \text{maximize:}\quad & \ip{W u u^{\ast}W^{\ast}}{Y}\\
      \text{subject to:}\quad 
      & Y \in \S_n(\Y_n,\ldots,\Y_1;\X_n,\ldots,\X_1)
    \end{aligned}
  \end{equation}
  have the same optimum value.
\end{cor}

\begin{remark}
  Using the notation introduced in \cite{ChiribellaE16}, which defines a
  quantum network generalization of conditional min-entropy, the equivalence
  expressed by Corollary~\ref{cor:main} may alternatively be written
  \begin{equation}
    \begin{multlined}
      \op{H}_{\text{min}}\bigl(\reg{Y}_n\mid\reg{X}_1,\reg{Y}_1,\ldots,
      \reg{X}_{n-1},\reg{Y}_{n-1},\reg{X}_n\bigr)_{u u^{\ast}}\\[1mm]
      = \op{H}_{\text{min}}\bigl(\reg{X}_1\mid\reg{Y}_n,\reg{X}_n,\ldots,
      \reg{Y}_{2},\reg{X}_{2},\reg{Y}_1\bigr)_{u u^{\ast}}
    \end{multlined}
  \end{equation}
  for every vector $u \in \X_1 \otimes\Y_1\otimes\cdots\otimes\X_n\otimes\Y_n$.
  \end{remark}

\subsection*{Interpretations of the main theorem}

Theorem~\ref{theorem:main} establishes a \emph{time-reversal property} of
rank-one strategy functions.
Intuitively speaking, the linear function
\begin{equation}
  Y \mapsto \ip{W u u^{\ast} W^{\ast}}{Y}
\end{equation}
defined on $\S_n(\Y_n,\ldots,\Y_1;\X_n,\ldots,\X_1)$ represents the
\emph{time-reversal} of the linear function
\begin{equation}
  X \mapsto \ip{u u^{\ast}}{X}
\end{equation}
defined on $\S_n(\X_1,\ldots,\X_n;\Y_1,\ldots,\Y_n)$, in the sense that the two
functions differ only in the reversal of the ordering of the register
exchanges:
$\reg{X}_1$, $\reg{Y}_1$, $\ldots$, $\reg{X}_n$, $\reg{Y}_n$ for the function
corresponding to $u u^{\ast}$ and $\reg{Y}_n$, $\reg{X}_n$, $\ldots$,
$\reg{Y}_1$, $\reg{X}_1$ for the function corresponding to
$W u u^{\ast} W^{\ast}$. 

For a given choice of $X\in\S_n(\X_1,\ldots,\X_n;\Y_1,\ldots,\Y_n)$, it is
generally not the case that
$W^{\ast} X W\in\S_n(\Y_n,\ldots,\Y_1;\X_n,\ldots,\X_1)$.
It may not even be the case that $W^{\ast} X W$ is the Choi representation of a
channel, and in the case that $W^{\ast} X W$ is the Choi representation of a
channel, it will generally not be the case that this channel obeys the
constraints necessary for it to be a valid strategy operator.
When combined with the observation that
$\S_n(\X_1,\ldots,\X_n;\Y_1,\ldots,\Y_n)$ and
$\S_n(\Y_n,\ldots,\Y_1;\X_n,\ldots,\X_1)$ are compact and convex sets, this
fact implies that the main theorem cannot possibly hold for all Hermitian
operators $H$ by the separating hyperplane theorem.
For small values of $n$ and for spaces having small dimensions, simple
examples of operators $H$ for which the main theorem fails may also easily be
obtained through random selections.

In Section~\ref{sec:entanglement-manipulation} we discuss another
interpretation of Theorem~\ref{theorem:main}, which concerns multiple round
entanglement manipulation.

\subsection*{Proof of Theorem~\ref{theorem:main}}

We will now prove Theorem~\ref{theorem:main}.
The first step is to express the strategy represented by $X$ as a sequence of
channels corresponding to invertible isometries (i.e., unitary operators for
which the input and output spaces have different names but necessarily the same
dimension), assuming an auxiliary input space initialized to a pure state
is made available.

\begin{figure}[t]
  \begin{center}
    \footnotesize
    \begin{tikzpicture}[scale=0.35, 
        turn/.style={draw, minimum height=14mm, minimum width=10mm,
          fill = ChannelColor, text=ChannelTextColor},
        smallturn/.style={draw, minimum width=9mm, minimum height=9mm,
          fill = ChannelColor},
        invisible/.style={minimum height=1mm, minimum width=10mm},
        >=latex]
      
      \node (U0) at (-16,4.4) [smallturn] {$v$};
      \node (U1) at (-8,4) [turn] {$U_1$};
      \node (U2) at (2,4) [turn] {$U_2$};
      \node (U3) at (12,4) [turn] {$U_3$};
      \node (U4) at (18.5,4) [invisible] {};
     
      \node (P0) at (-13,-2) [invisible] {};
      \node (P1) at (-3,-2) [invisible] {};
      \node (P2) at (7,-2) [invisible] {};
      \node (P3) at (17,-2) [invisible] {};

      \draw[->] (U0.east) -- ([yshift=4mm]U1.west)
      node [above, midway] {$\reg{Z}_0$};

      \draw[->] ([yshift=4mm]U1.east) -- ([yshift=4mm]U2.west)
      node [above, midway] {$\reg{Z}_1$};
      
      \draw[->] ([yshift=4mm]U2.east) -- ([yshift=4mm]U3.west)
      node [above, midway] {$\reg{Z}_2$};
      
      \draw[->] ([yshift=4mm]U3.east) -- ([yshift=4mm]U4.west)
      node [above, midway] {$\reg{Z}_3$};
      
      \draw[->] ([yshift=-4mm]U1.east) .. controls +(right:20mm) and 
      +(left:20mm) .. ([yshift=4mm]P1.west) node [right, pos=0.5] {$\reg{Y}_1$};
      
      \draw[->] ([yshift=-4mm]U2.east) .. controls +(right:20mm) and 
      +(left:20mm) .. ([yshift=4mm]P2.west) node [right, pos=0.5] {$\reg{Y}_2$};
      
      \draw[->] ([yshift=-4mm]U3.east) .. controls +(right:20mm) and 
      +(left:20mm) .. ([yshift=4mm]P3.west) node [right, pos=0.5] {$\reg{Y}_3$};
      
      \draw[->] ([yshift=4mm]P0.east) .. controls +(right:20mm) and 
      +(left:20mm) .. ([yshift=-4mm]U1.west) node [left, pos=0.5] {$\reg{X}_1$};
      
      \draw[->] ([yshift=4mm]P1.east) .. controls +(right:20mm) and 
      +(left:20mm) .. ([yshift=-4mm]U2.west) node [left, pos=0.5] {$\reg{X}_2$};
      
      \draw[->] ([yshift=4mm]P2.east) .. controls +(right:20mm) and 
      +(left:20mm) .. ([yshift=-4mm]U3.west) node [left, pos=0.5] {$\reg{X}_3$};
      
    \end{tikzpicture}\\[6mm]
    \begin{tikzpicture}[scale=0.35, 
        turn/.style={draw, minimum height=14mm, minimum width=10mm,
          fill = ChannelColor, text=ChannelTextColor},
        smallturn/.style={draw, minimum width=9mm, minimum height=9mm,
          fill = ChannelColor},
        invisible/.style={minimum height=1mm, minimum width=10mm},
        >=latex]
      
      \node (U0) at (-16,4.4) [smallturn] {$w$};
      \node (U1) at (-8,4) [turn] {$U_3^{\t}$};
      \node (U2) at (2,4) [turn] {$U_2^{\t}$};
      \node (U3) at (12,4) [turn] {$U_1^{\t}$};
      \node (U4) at (18.5,4) [invisible] {};
     
      \node (P0) at (-13,-2) [invisible] {};
      \node (P1) at (-3,-2) [invisible] {};
      \node (P2) at (7,-2) [invisible] {};
      \node (P3) at (17,-2) [invisible] {};

      \draw[->] (U0.east) -- ([yshift=4mm]U1.west)
      node [above, midway] {$\reg{Z}_3$};

      \draw[->] ([yshift=4mm]U1.east) -- ([yshift=4mm]U2.west)
      node [above, midway] {$\reg{Z}_2$};
      
      \draw[->] ([yshift=4mm]U2.east) -- ([yshift=4mm]U3.west)
      node [above, midway] {$\reg{Z}_1$};
      
      \draw[->] ([yshift=4mm]U3.east) -- ([yshift=4mm]U4.west)
      node [above, midway] {$\reg{Z}_0$};
      
      \draw[->] ([yshift=-4mm]U1.east) .. controls +(right:20mm) and 
      +(left:20mm) .. ([yshift=4mm]P1.west) node [right, pos=0.5] {$\reg{X}_3$};
      
      \draw[->] ([yshift=-4mm]U2.east) .. controls +(right:20mm) and 
      +(left:20mm) .. ([yshift=4mm]P2.west) node [right, pos=0.5] {$\reg{X}_2$};
      
      \draw[->] ([yshift=-4mm]U3.east) .. controls +(right:20mm) and 
      +(left:20mm) .. ([yshift=4mm]P3.west) node [right, pos=0.5] {$\reg{X}_1$};
      
      \draw[->] ([yshift=4mm]P0.east) .. controls +(right:20mm) and 
      +(left:20mm) .. ([yshift=-4mm]U1.west) node [left, pos=0.5] {$\reg{Y}_3$};
      
      \draw[->] ([yshift=4mm]P1.east) .. controls +(right:20mm) and 
      +(left:20mm) .. ([yshift=-4mm]U2.west) node [left, pos=0.5] {$\reg{Y}_2$};
      
      \draw[->] ([yshift=4mm]P2.east) .. controls +(right:20mm) and 
      +(left:20mm) .. ([yshift=-4mm]U3.west) node [left, pos=0.5] {$\reg{Y}_1$};
      
    \end{tikzpicture}
  \end{center}
  \caption{An arbitrary strategy may be implemented by initializing a register
    $\reg{Z}_0$ to a pure state $v$, followed by the application of an
    invertible isometric channel on each turn, and finally by discarding the
    last memory register $\reg{Z}_n$ (which is $\reg{Z}_3$ in the picture).
    The time-reversed strategy whose existence is implied by the main theorem
    is obtained by setting the register $\reg{Z}_n$ ($\reg{Z}_3$ in the
    picture) to an appropriate choice of a pure state $w$, followed by
    the application of invertible isometric channels obtained by transposing
    the original isometries, and finally by discarding the memory register
    $\reg{Z}_0$.}
  \label{fig:unitary-Alice}
\end{figure}
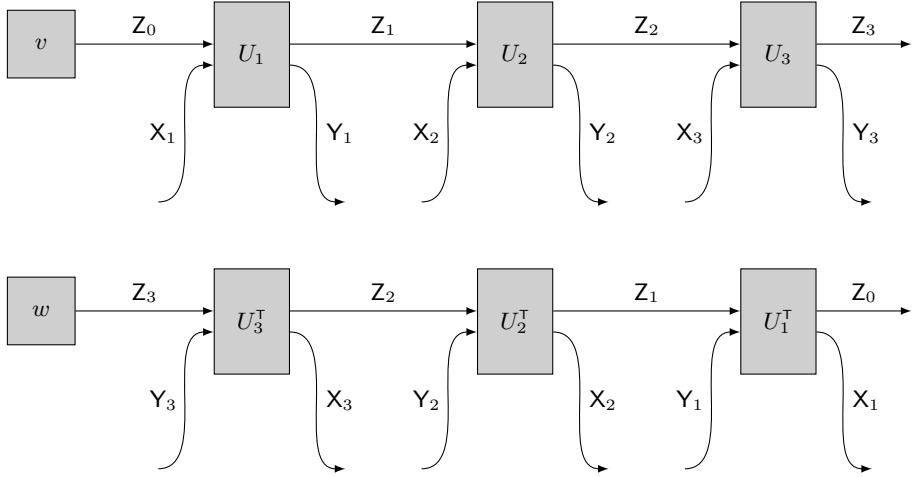

Through the repeated application of the Stinespring dilation theorem, together
with the result of \cite{GutoskiW07,ChiribellaDP08,ChiribellaDP09} establishing
that $X=J(\Xi_n)$ is the Choi representation of a channel $\Xi_n$ arising from
a valid $n$-turn strategy, one finds that there must exist complex Euclidean
spaces $\Z_0,\ldots,\Z_n$ satisfying
$\dim(\Z_{k-1}\otimes\X_k) = \dim(\Z_k\otimes\Y_k)$ for all
$k\in\{1,\ldots,n\}$, a unit vector $v \in \Z_0$, and invertible isometries
$U_1,\ldots,U_n$ of the form
\begin{equation}
  U_k\in\Unitary(\Z_{k-1}\otimes \X_k,\Y_k\otimes\Z_k)
\end{equation}
such that
\begin{equation}
  \label{eq:Xi_n}
  \Xi_n(Z) = \tr_{\Z_n} \bigl(U (v v^{\ast} \otimes Z\bigr) U^{\ast}\bigr)
\end{equation}
for all $Z\in\Lin(\X_1\otimes\cdots\otimes\X_n)$, where
\begin{equation}
  \begin{multlined}
    U = (\I_{\Y_1\otimes\cdots\otimes\Y_{n-1}}\otimes U_n)
    \cdots (U_1 \otimes \I_{\X_2\otimes\cdots\otimes\X_n})\\[1mm]
    \in\Unitary(\Z_0\otimes\X_1\otimes\cdots\otimes\X_n,
    \Y_1\otimes\cdots\otimes\Y_n\otimes\Z_n).
  \end{multlined}
\end{equation}
In words, the strategy represented by the operator $X$ is implemented by
first initializing a register $\reg{Z}_0$ to the pure state $v$, then
applying the invertible isometric channels corresponding to $U_1,\ldots,U_n$,
and finally discarding $\reg{Z}_n$ after the interaction has finished.
(The top picture in Figure~\ref{fig:unitary-Alice} illustrates this for the
case $n=3$.)

The vector $u$ may be expressed as
\begin{equation}
  u = \sum_{\substack{a_1,\ldots,a_n\\b_1,\ldots,b_n}}
  u(b_1,\ldots,b_n,a_1,\ldots,a_n)
  \ket{b_1}\cdots\ket{b_n}\ket{a_1}\cdots\ket{a_n},
\end{equation}
where the sum is over all standard basis states $\ket{a_1},\ldots,\ket{a_n}$ of
$\X_1,\ldots,\X_n$ and $\ket{b_1},\ldots,\ket{b_n}$ of $\Y_1,\ldots,\Y_n$,
respectively.
Based on this expression, define an operator $A\in\Lin(\Z_0,\Z_n)$ as
\begin{equation}
  \begin{multlined}
    A = \sum_{\substack{a_1,\ldots,a_n\\b_1,\ldots,b_n}}
    u(b_1,\ldots,b_n,a_1,\ldots,a_n)
    (\bra{b_n}\otimes\I_{\Z_n})U_n(\I_{\Z_{n-1}}\otimes\ket{a_n})
    \rule{10mm}{0mm}\\[-4mm]
    \cdots
    (\bra{b_1}\otimes\I_{\Z_1})U_1(\I_{\Z_{0}}\otimes\ket{a_1}).
  \end{multlined}
\end{equation}
By considering the action of the strategy represented by $v$ and
$U_1,\ldots,U_n$, then performing the required operator-vector multiplications
required to evaluate the expression $\ip{u u^{\ast}}{X}$
when $X = J(\Xi_n)$ for $\Xi_n$ given by \eqref{eq:Xi_n}, one concludes that
\begin{equation}
  \ip{u u^{\ast}}{X} = \norm{A v}^2 = \ip{v v^{\ast}}{A^{\ast} A}.
\end{equation}

Next we turn to the reversed interaction.
To obtain a strategy operator $Y$ satisfying the requirements of the theorem,
we consider the strategy obtained by initializing the register $\reg{Z}_n$ to
a particular choice of a pure state $w$, which will be selected later,
then applying in sequence the invertible isometric channels corresponding to
the operators $U_n^{\t},\ldots,U_1^{\t}$.
(The bottom picture in Figure~\ref{fig:unitary-Alice} illustrates this for the
case $n=3$.)
That is, for 
\begin{equation}
  \begin{multlined}
    V = (\I_{\X_n\otimes\cdots\otimes\X_{2}}\otimes U_1^{\t})
    \cdots (U_n^{\t} \otimes \I_{\Y_{n-1}\otimes\cdots\otimes\Y_1})\\[1mm]
    \in\Unitary(\Z_n\otimes\Y_n\otimes\cdots\otimes\Y_1,
    \Z_0\otimes\X_n\otimes\cdots\otimes\X_1),
  \end{multlined}
\end{equation}
we consider the channel $\Theta_n\in\Channel(\Y_n\otimes\cdots\otimes\Y_1,
\X_n\otimes\cdots\otimes\X_1)$ defined as
\begin{equation}
  \Theta_n(Z) = \tr_{\Z_0}\bigl( V (w w^{\ast} \otimes Z) V^{\ast}\bigr)
\end{equation}
for all $Z\in\Lin(\Y_n\otimes\cdots\otimes\Y_1)$.
It is evident from the specification of this channel,
irrespective of the choice of the pure state $w$, that
$Y = J(\Theta_n)\in\S_n(\Y_n,\ldots,\Y_1;\X_n,\ldots,\X_1)$.
By considering the action of this strategy, a similar calculation to the one
above reveals that 
\begin{equation}
  \ip{Wu u^{\ast}W^{\ast}}{Y} = \norm{A^{\t} w}^2 =
  \bigip{w w^{\ast}}{\overline{A}A^{\t}}.
\end{equation}

The nonzero eigenvalues of $A^{\ast} A$ and $\overline{A} A^{\t}$ are equal,
and therefore by choosing $w$ to be an eigenvector corresponding to the largest
eigenvalue of $\overline{A} A^{\t}$ one obtains
\begin{equation}
  \label{eq:final-inequality}
  \ip{Wu u^{\ast}W^{\ast}}{Y} =
  \bigip{w w^{\ast}}{\overline{A}A^{\t}}
  \geq \ip{v v^{\ast}}{A^{\ast} A} = \ip{u u^{\ast}}{X}.
\end{equation}
If it holds that $\dim(\Y_1\otimes\cdots\otimes\Y_n)\leq
\dim(\X_1\otimes\cdots\otimes\X_n)$, then $\dim(\Z_0)\leq\dim(\Z_n)$, which
implies that the inequality in \eqref{eq:final-inequality} may be taken as an
equality for an appropriate choice of a pure state $w$.
This completes the proof.

\section{Application to min- and max-entropy}

In this section we connect the main result proved in the previous section to
the conditional min- and max-entropy functions.
These function, which were first introduced in \cite{Datta09}, may be defined
as follows.
First, one defines the max- and min-relative entropy of $P$ with respect to
$Q$, for positive semidefinite operators $P$ and $Q$ (acting on the same
space), as follows:
\begin{align}
  \op{D}_{\text{max}}(P \,\|\, Q) & = \log\bigl(
  \min\{\lambda \geq 0\,:\, P \leq \lambda Q\}\bigr),\\
  \op{D}_{\text{min}}(P \,\|\, Q) & = -\log\bigl(\fid(P,Q)^2\bigr).
\end{align}
Then, with respect to a given state $\rho\in\Density(\X\otimes\Y)$ of a pair of
registers $(\reg{X},\reg{Y})$, one defines
\begin{align}
  \op{H}_{\text{min}}(\reg{X} | \reg{Y}) = -\inf_{\sigma\in\Density(\Y)}
  \op{D}_{\text{max}}\bigl(\rho\,\big\|\,\I_{\X} \otimes \sigma\bigr),\\
  \op{H}_{\text{max}}(\reg{X} | \reg{Y}) = -\inf_{\sigma\in\Density(\Y)}
  \op{D}_{\text{min}}\bigl(\rho\,\big\|\,\I_{\X} \otimes \sigma\bigr).
\end{align}
It is known that these two quantities are related in the following way:
with respect to any pure state $u u^{\ast}$ of a triple of registers
$(\reg{X},\reg{Y},\reg{Z})$, one has that
\begin{equation}
  \label{eq:relation-min-and-max-entropy}
  \op{H}_{\text{min}}(\reg{X} | \reg{Y})
  = -\op{H}_{\text{max}}(\reg{X} | \reg{Z}).
\end{equation}
(Indeed, in \cite{KoenigRS09} the conditional max-relative entropy of a
state of $(\reg{X},\reg{Z})$ is \emph{defined} by the equation
\eqref{eq:relation-min-and-max-entropy}, which does not depend on which
purification of this state is chosen, and is then proved to agree with the
definition stated previously.)

Consider any unit vector $u \in \X \otimes \Y \otimes \Z$, which defines a pure
state $u u^{\ast}$ of a triple of registers $(\reg{X},\reg{Y},\reg{Z})$.
We will consider two optimization problems defined by $u$, the first of which
is as follows:
\begin{equation}
  \label{eq:min-entropy-problem}
  \begin{aligned}
    \text{maximize:} &  \quad\bigip{ u u^{\ast} }{ X } \\
    \text{subject to:} & \quad X \in \S_2(\Y,\Z;\X,\complex).
  \end{aligned}
\end{equation}
This optimization problem is illustrated in Figure~\ref{fig:min-entropy}.
\begin{figure}[t]
  \begin{center}
    \begin{tikzpicture}[scale=0.4, 
        turn/.style={draw, minimum height=14mm, minimum width=10mm,
          fill = ChannelColor, text=ChannelTextColor},
        measure/.style={draw, minimum width=7mm, minimum height=7mm,
          fill = ChannelColor},
        >=latex]
      
      \node (B1) at (-10,-4) [minimum height=6mm] {};
      \node (B2) at (0,-4) [minimum height=6mm] {};
      \node (B3) at (3,-4) [minimum height=6mm] {};

      \node (B) at (-3.5,-4) [draw, fill=ChannelColor, minimum height=6mm,
        minimum width = 6cm] {$u u^{\ast}$};
      
      \node (A1) at (-5,2) [turn] {$\Phi_1$};
      \node (A2) at (8,2) [turn] {$\Phi_2$};

      \draw[->] (B1.north) .. controls +(up:20mm) and 
      +(left:20mm) .. ([yshift=-4mm]A1.west) node [above left, pos=0.5]
      {$\reg{Y}$};
      
      \draw[->] (B3.north) .. controls +(up:20mm) and 
      +(left:20mm) .. ([yshift=-4mm]A2.west) node [above left, pos=0.5]
      {$\reg{Z}$};
      
      \draw[->] ([yshift=-4mm]A1.east) .. controls +(right:20mm) and 
      +(up:20mm) .. (B2.north) node [above right, pos=0.5] {$\reg{X}$};
      
      \draw[->] ([yshift=4mm]A1.east) -- ([yshift=4mm]A2.west)
      node [above, midway] {$\reg{W}$};
      
    \end{tikzpicture}
  \end{center}
  \caption{The optimization problem \eqref{eq:min-entropy-problem} corresponds
    to a maximization of the linear functions defined by $u u^{\ast}$ over all
    strategies given by channels $\Phi_1$ and $\Phi_2$, for an arbitrary 
    choice of a register $\reg{W}$.}
  \label{fig:min-entropy}
\end{figure}
In this case, the channel $\Phi_2$ takes registers $\reg{Z}$ and $\reg{W}$
as input and outputs nothing (which is equivalent to outputting the unique
state $1 \in \Density(\complex)$ of a one-dimensional system).
That is, $\Phi_2$ must be the trace mapping.
One may therefore simplify this problem, obtaining the following semidefinite
program:
\begin{center}
  \begin{minipage}[t]{.42\textwidth}
    \centerline{\underline{Primal problem}}\vspace{-6mm}
    \begin{align*}
      \text{maximize:}\quad & \ip{\tr_{\Z}(u u^{\ast})}{X}\\
      \text{subject to:}\quad 
      & \tr_{\X}(X) = \I_{\Y},\\
      & X \in \Pos(\X\otimes\Y).
    \end{align*}
  \end{minipage}
  \begin{minipage}[t]{.42\textwidth}
    \centerline{\underline{Dual problem}}\vspace{-6mm}
    \begin{align*}
      \text{minimize:}\quad & \tr(Y)\\
      \text{subject to:}\quad 
      & \I_{\X} \otimes Y \geq \tr_{\Z}(u u^{\ast}),\\
      & Y \in \Herm(\X).
    \end{align*}
  \end{minipage}
\end{center}
By examining the dual problem, one sees that the optimal value of this
semidefinite program is
\begin{equation}
  \label{eq:H_min-SDP-optimum}
  2^{-\op{H}_{\text{min}}(\reg{X}|\reg{Y})}
\end{equation}
with respect to the state $u u^{\ast}$ of $(\reg{X},\reg{Y},\reg{Z})$.
K\"onig, Renner, and Schaffner \cite{KoenigRS09} observed that the primal
problem coincides with the value represented by the expression
\eqref{eq:H_min-SDP-optimum}, which is consistent with the observation that
strong duality always holds for this semidefinite program (which may be
verified through Slater's theorem, for instance).

\begin{figure}[t]
  \begin{center}
    \begin{tikzpicture}[scale=0.4, 
        turn/.style={draw, minimum height=14mm, minimum width=10mm,
          fill = ChannelColor, text=ChannelTextColor},
        measure/.style={draw, minimum width=7mm, minimum height=7mm,
          fill = ChannelColor},
        >=latex]
      
      \node (B1) at (0,-4) [minimum height=6mm] {};
      \node (B2) at (3,-4) [minimum height=6mm] {};
      \node (B3) at (13,-4) [minimum height=6mm] {};

      \node (B) at (6.5,-4) [draw, fill=ChannelColor, minimum height=6mm,
        minimum width = 6cm] {$W u u^{\ast} W^{\ast}$};
      
      \node (A1) at (-5,2) [turn] {$\Psi_1$};
      \node (A2) at (8,2) [turn] {$\Psi_2$};
      
      \draw[->] (B2.north) .. controls +(up:20mm) and 
      +(left:20mm) .. ([yshift=-4mm]A2.west) node [above left, pos=0.5]
      {$\reg{X}$};
      
      \draw[->] ([yshift=-4mm]A1.east) .. controls +(right:20mm) and 
      +(up:20mm) .. (B1.north) node [above right, pos=0.5] {$\reg{Z}$};
      
      \draw[->] ([yshift=-4mm]A2.east) .. controls +(right:20mm) and 
      +(up:20mm) .. (B3.north) node [above right, pos=0.5] {$\reg{Y}$};
      
      \draw[->] ([yshift=4mm]A1.east) -- ([yshift=4mm]A2.west)
      node [above, midway] {$\reg{W}$};
      
    \end{tikzpicture}
  \end{center}
  \caption{The optimization problem \eqref{eq:max-entropy-problem} corresponds
    to a maximization of the linear functions defined by $Wuu^{\ast}W^{\ast}$
    over all strategies given by channels $\Psi_1$ and $\Psi_2$, for an
    arbitrary choice of a register $\reg{W}$.}
  \label{fig:max-entropy}
\end{figure}
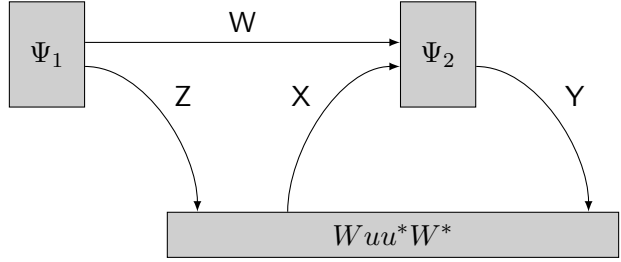

The second optimization problem we consider is the time-reversal of the first,
and may be stated as follows:
\begin{equation}
  \label{eq:max-entropy-problem}
  \begin{aligned}
    \text{maximize:} &  \quad\bigip{ W u u^{\ast} W^{\ast} }{ Y } \\
    \text{subject to:} & \quad Y \in \S_2(\complex,\X;\Z,\Y).
  \end{aligned}
\end{equation}
Figure~\ref{fig:max-entropy} illustrates the interaction corresponding to this
optimization problem.
The inclusion $X\in\S_2(\complex,\X;\Z,\Y)$, for a given operator
$X\in\Pos(\Z\otimes\Y\otimes\X)$, is equivalent to the condition that
$\tr_{\Y}(X) = \sigma\otimes\I_{\X}$ for some $\sigma\in\Density(\Z)$.
After re-ordering tensor factors, we obtain the following semidefinite program:
\begin{center}
  \begin{minipage}[t]{.42\textwidth}
    \centerline{\underline{Primal problem}}\vspace{-6mm}
    \begin{align*}
      \text{maximize:}\quad & \ip{u u^{\ast}}{X}\\
      \text{subject to:}\quad 
      & \tr_{\Y}(X) = \I_{\X} \otimes \sigma,\\
      & X \in \Pos(\X\otimes\Y\otimes\Z),\\
      & \sigma \in \Density(\Z).
    \end{align*}
  \end{minipage}
  \begin{minipage}[t]{.42\textwidth}
    \centerline{\underline{Dual problem}}\vspace{-6mm}
    \begin{align*}
      \text{minimize:}\quad & \lambda\\
      \text{subject to:}\quad 
      & Y \otimes \I_{\Y} \geq u u^{\ast},\\
      & \lambda \I_{\Z} \geq \tr_{\X}(Y),\\
      & Y \in \Herm(\X\otimes\Z),\\
      & \lambda \in \real.
    \end{align*}
  \end{minipage}
\end{center}
An examination of the primal problem reveals (through Uhlmann's theorem) that
the optimal value of this semidefinite program is
\begin{equation}
  \label{eq:H_max-SDP-optimum}
  2^{\op{H}_{\text{max}}(\reg{X}|\reg{Z})}.
\end{equation}

By our main theorem, it follows that the two optimization problems have the
same optimal value, and therefore we obtain an alternative proof that with
respect to every pure state of a triple or registers
$(\reg{X},\reg{Y},\reg{Z})$ one has
\begin{equation}
  \label{eq:min-max-entropy-identity}
  \op{H}_{\text{min}}(\reg{X} | \reg{Y})
  = -\op{H}_{\text{max}}(\reg{X} | \reg{Z}).
\end{equation}

\begin{figure}[t]
  \begin{center}
    \begin{tikzpicture}[scale=0.4, 
        turn/.style={draw, minimum height=14mm, minimum width=10mm,
          fill = ChannelColor, text=ChannelTextColor},
        >=latex]
      
      \node (B1) at (-10,-4) [minimum height=6mm] {};
      \node (B2) at (0,-4) [minimum height=6mm] {};
      \node (B3) at (3,-4) [minimum height=6mm] {};
      \node (B4) at (13,-4) [minimum height=6mm] {};

      \node (B) at (1.5,-4) [draw, fill=ChannelColor, minimum height=6mm,
        minimum width = 10cm] {$u u^{\ast}$};
      
      \node (A1) at (-5,2) [turn] {$\Phi_1$};
      \node (A2) at (8,2) [turn] {$\Phi_2$};

      \draw[->] (B1.north) .. controls +(up:20mm) and 
      +(left:20mm) .. ([yshift=-4mm]A1.west) node [above left, pos=0.5]
      {$\reg{Y}$};
      
      \draw[->] (B3.north) .. controls +(up:20mm) and 
      +(left:20mm) .. ([yshift=-4mm]A2.west) node [above left, pos=0.5]
      {$\reg{Z}$};
      
      \draw[->] ([yshift=-4mm]A1.east) .. controls +(right:20mm) and 
      +(up:20mm) .. (B2.north) node [above right, pos=0.5] {$\reg{X}$};
      
      \draw[->] ([yshift=-4mm]A2.east) .. controls +(right:20mm) and 
      +(up:20mm) .. (B4.north) node [above right, pos=0.5] {$\reg{W}$};
      
      \draw[->] ([yshift=4mm]A1.east) -- ([yshift=4mm]A2.west);
      
    \end{tikzpicture}
  \end{center}
  \caption{
    Maximizing the linear function defined by $u u^{\ast}$ over all
    four-message strategies of the form depicted yields the left-hand side of
    \eqref{eq:four-message-identity}.
    By reversing time, the right-hand side of that equation is obtained,
    and the equality of the two is implied by the main theorem.}
  \label{fig:four-message-interaction}
\end{figure}
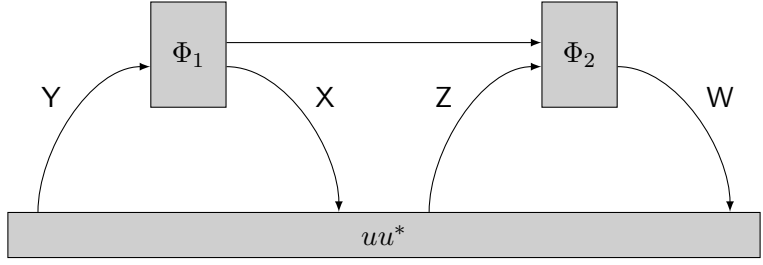

It is natural to ask if the connections among min-entropy, max-entropy, and
optimization problems involving three-message strategies have interesting
implications or generalizations for interactions involving four or more
messages.
As a partial answer to this question, we observe that when our main result is
applied to the four-message interaction depicted in
Figure~\ref{fig:four-message-interaction}, it reveals the identity
\begin{equation}
  \label{eq:four-message-identity}
  \max_{\Phi\in\Channel(\Y,\X)} \fid\bigl(
  \tr_{\W}(u u^{\ast}), J(\Phi) \otimes \I_{\Z}\bigr)
  = \max_{\Psi\in\Channel(\W,\Z)} \fid\bigl(
  \tr_{\X}(u u^{\ast}), \I_{\Y} \otimes J(\Psi)\bigr)
\end{equation}
for all vectors $u\in\X\otimes\Y\otimes\Z\otimes\W$.
This identity is appealing in its simplicity and symmetry, and by taking
$\W=\complex$ (or $\Y=\complex$) a statement equivalent to
\eqref{eq:min-max-entropy-identity} for all pure states of
$(\reg{X},\reg{Y},\reg{Z})$ is obtained.
We do not know, however, if the quantity represented by either side of the
identity has any direct operational significance.

Other identities may be obtained through a similar methodology, although they
become increasingly complex as the number of messages is increased.

\section{Online pure state entanglement manipulation}
\label{sec:entanglement-manipulation}

The following three statements are equivalent for a given operator
$X\in\Lin(\Y\otimes\X)$:
\begin{enumerate}
\item[1.]
  $X\in\S_1(\X;\Y)$.
  (Equivalently, $X\in\Pos(\Y\otimes\X)$ and $\tr_{\Y}(X) = \I_{\X}$.)
\item[2.] $X = (\Phi\otimes\I_{\Lin(\X)})(\vec(\I_{\X})\vec(\I_{\X})^{\ast})$
  for some channel $\Phi\in\Channel(\X,\Y)$.
\item[3.] $X = (\I_{\Lin(\Y)}\otimes \Psi)(\vec(\I_{\Y})\vec(\I_{\Y})^{\ast})$
  for some completely positive and unital map $\Psi\in\CP(\Y,\X)$.
\end{enumerate}
(Here and throughout this section, $\text{vec}$ refers to the vectorization
mapping, which is the mapping obtained by extending the transformation
$\ket{a}\bra{b} \mapsto \ket{a}\ket{b}$ for standard basis states to arbitrary
operators by linearity.
In particular, $\vec(\I_{\X})$ is a non-normalized vector proportional to the
canonical maximally entangled pure state corresponding to two identical copies
of a system whose state space is $\X$.)
The maps $\Phi$ and $\Psi$ uniquely determine one another, and it is reasonable
to view these maps as being related by transposition (with respect to the
standard basis): $\Psi = \Phi^{\t}$ and $\Phi = \Psi^{\t}$.
To obtain a Kraus representation for $\Psi$, for instance, one may simply
take a Kraus representation of $\Phi$ and transpose each of the Kraus
operators.
(The transpose of an arbitrary map can be defined in a manner that is
consistent with these statements, but it is sufficient for our needs to focus
on channels and completely positive unital maps.)

A generalization of the equivalence mentioned above to the quantum strategy
framework may also be verified.
For an operator
$X\in\Lin(\Y_1\otimes\cdots\otimes\Y_n\otimes\X_1\otimes\cdots\otimes\X_n)$,
these three statements are equivalent:
\begin{enumerate}
\item[1.] $X\in\S_n(\X_1,\ldots,\X_n;\Y_1,\ldots,\Y_n)$.
\item[2.] There exist complex Euclidean spaces $\Z_1,\ldots,\Z_{n-1}$
  (and $\Z_0 = \complex$ and $\Z_n = \complex$), along
  with channels $\Phi_1, \ldots,\Phi_n$ having the form
  \begin{equation}
    \Phi_k \in \Channel(\Z_{k-1}\otimes\X_k,\Y_k\otimes\Z_k),
  \end{equation}
  such that the channel
  $\Xi_n\in\Channel(\X_1\otimes\cdots\otimes\X_n,\Y_1\otimes\cdots\otimes\Y_n)$
  defined as
  \begin{equation}
    \Xi_n = \bigl(\I_{\Lin(\Y_1\otimes\cdots\otimes\Y_{n-1})}\otimes\Phi_n\bigr)
    \cdots \bigl(\Phi_1 \otimes \I_{\Lin(\X_2\otimes\cdots\otimes\X_n)}\bigr)
  \end{equation}
  satisfies
  \begin{equation}
    X = \bigl(\Xi_n \otimes \I_{\Lin(\X_1\otimes\cdots\otimes\X_n)}\bigr)
    (\vec(\I_{\X_1\otimes\cdots\otimes\X_n})
    \vec(\I_{\X_1\otimes\cdots\otimes\X_n})^{\ast}).
  \end{equation}
\item[3.] There exist complex Euclidean spaces $\Z_1,\ldots,\Z_{n-1}$
  (and $\Z_0 = \complex$ and $\Z_n = \complex$), along
  with completely positive and unital maps $\Psi_1, \ldots,\Psi_n$ having the
  form
  \begin{equation}
    \Psi_k \in \Channel(\Y_k\otimes\Z_{k},\Z_{k-1}\otimes\X_k),
  \end{equation}
  such that the unital map
  $\Lambda_n\in\CP(\Y_1\otimes\cdots\otimes\Y_n,\X_1\otimes\cdots\otimes\X_n)$
  defined as
  \begin{equation}
    \Lambda_n =
    \bigl(\Psi_1 \otimes \I_{\Lin(\X_2\otimes\cdots\otimes\X_n)}\bigr)\cdots
    \bigl(\I_{\Lin(\Y_1\otimes\cdots\otimes\Y_{n-1})}\otimes\Psi_n\bigr)
  \end{equation}
  satisfies
  \begin{equation}
    X = \bigl(\I_{\Lin(\Y_1\otimes\cdots\otimes\Y_n)}\otimes\Lambda_n\bigr)
    (\vec(\I_{\Y_1\otimes\cdots\otimes\Y_n})
    \vec(\I_{\Y_1\otimes\cdots\otimes\Y_n})^{\ast}).
  \end{equation}
\end{enumerate}

Through this equivalence, for a given state
$\rho\in\Density(\Y_1\otimes\cdots\otimes\Y_n
\otimes\X_1\otimes\cdots\otimes\X_n)$,
one arrives at an alternative interpretation of the semidefinite program
\begin{equation}
  \label{eq:SDP-entanglement-concentration}
  \begin{aligned}
    \text{maximize:} \quad & \ip{\rho}{X}\\
    \text{subject to:} \quad & X\in
    \S_n(\X_1,\ldots,\X_n;\Y_1,\ldots,\Y_n)
  \end{aligned}
\end{equation}
that concerns an online variant of entanglement manipulation, as is explained
shortly.
The term ``online'' in this context refers to a situation in which a quantum
state must be manipulated in multiple turns, where an output is required
immediately after each input system arrives and prior to the next input system
being made available, similar to an online process.

By the equivalence of the third statement above to the first, a maximization
over all $X\in\S_n(\X_1,\ldots,\X_n;\Y_1,\ldots,\Y_n)$ is equivalent to a
maximization over all operators
\begin{equation}
  \bigl(\I_{\Lin(\Y_1\otimes\cdots\otimes\Y_n)}\otimes\Lambda_n\bigr)
  (\vec(\I_{\Y_1\otimes\cdots\otimes\Y_n})
  \vec(\I_{\Y_1\otimes\cdots\otimes\Y_n})^{\ast})
\end{equation}
for 
\begin{equation}
    \Lambda_n =
    \bigl(\Psi_1 \otimes \I_{\Lin(\X_2\otimes\cdots\otimes\X_n)}\bigr)\cdots
    \bigl(\I_{\Lin(\Y_1\otimes\cdots\otimes\Y_{n-1})}\otimes\Psi_n\bigr)
  \end{equation}
and $\Psi_1,\ldots,\Psi_n$ being completely positive and unital maps of
the form
\begin{equation}
  \Psi_k \in \Channel(\Y_k\otimes\Z_k,\Z_{k-1}\otimes\X_k).
\end{equation}
The value of the objective function $\ip{\rho}{X}$ may therefore be expressed
as
\begin{equation}
  \bigip{
    (\I_{\Lin(\Y_1\otimes\cdots\otimes\Y_n)}\otimes\Lambda_n^{\ast})(\rho)
  }{
    \vec(\I_{\Y_1\otimes\cdots\otimes\Y_n})
    \vec(\I_{\Y_1\otimes\cdots\otimes\Y_n})^{\ast}
  },
\end{equation}
which is $\dim(\Y_1\otimes\cdots\otimes\Y_n)$ times the squared fidelity
between the maximally entangled state
$\tau\in\Density(\Y_1\otimes\cdots\otimes\Y_n\otimes
\Y_1\otimes\cdots\otimes\Y_n)$ given by
\begin{equation}
  \tau = \frac{\vec(\I_{\Y_1\otimes\cdots\otimes\Y_n})
       \vec(\I_{\Y_1\otimes\cdots\otimes\Y_n})^{\ast}}{
    \dim(\Y_1\otimes\cdots\otimes\Y_n)}
\end{equation}
and the state obtained by applying the channel $\Lambda_n^{\ast}$ to the
portion of $\rho$ corresponding to the spaces $\X_1,\ldots,\X_n$.
In the case that $n=1$, K\"onig, Renner, and Schaffner \cite{KoenigRS09}
refer to this quantity as the \emph{quantum correlation}.
This situation is illustrated for the case $n=3$ in
Figure~\ref{fig:online-entanglement-manipulation-1}.

\begin{figure}[p]
  \begin{center}
    \begin{tikzpicture}[scale=0.4, 
        turn/.style={draw, minimum height=12mm, minimum width=10mm,
          fill = ChannelColor, text=ChannelTextColor},
        state/.style={draw, minimum width=10mm, minimum height=70mm,
          fill = ChannelColor},
        invisible/.style={minimum height=1mm, minimum width=10mm},
        >=latex]

      \node (rho) at (-12,-4.5) [state] {$\rho$};

      \node (X1) at (-12,-6) [invisible] {};
      \node (X2) at (-12,-9) [invisible] {};
      \node (X3) at (-12,-12) [invisible] {};
      
      \node (Y1) at (12,-6) [invisible] {};
      \node (Y2) at (12,-9) [invisible] {};
      \node (Y3) at (12,-12) [invisible] {};
      
      \node (leftZ1) at (-12,3) [invisible] {};
      \node (leftZ2) at (-12,0) [invisible] {};
      \node (leftZ3) at (-12,-3) [invisible] {};
      
      \node (rightZ1) at (12,3) [invisible] {};
      \node (rightZ2) at (12,0) [invisible] {};
      \node (rightZ3) at (12,-3) [invisible] {};
     
      \node (B1) at (-6,-6) [turn] {$\Psi_1^{\ast}$};
      \node (B2) at (0,-9) [turn] {$\Psi_2^{\ast}$};
      \node (B3) at (6,-12) [turn] {$\Psi_3^{\ast}$};

      \draw[->] ([yshift=-3mm]X1.east) -- ([yshift=-3mm]B1.west)
      node [above, xshift=5mm, pos=0] {$\reg{X}_1$};

      \draw[->] ([yshift=-3mm]X2.east) -- ([yshift=-3mm]B2.west)
      node [above, xshift=5mm, pos=0] {$\reg{X}_2$};

      \draw[->] ([yshift=-3mm]X3.east) -- ([yshift=-3mm]B3.west)
      node [above, xshift=5mm, pos=0] {$\reg{X}_3$};

      \draw[->] ([yshift=3mm]B1.east) -- ([yshift=3mm]Y1.west)
      node [above, xshift=-5mm, pos=1] {$\reg{Y}_1$};

      \draw[->] ([yshift=3mm]B2.east) -- ([yshift=3mm]Y2.west)
      node [above, xshift=-5mm, pos=1] {$\reg{Y}_2$};

      \draw[->] ([yshift=3mm]B3.east) -- ([yshift=3mm]Y3.west)
      node [above, xshift=-5mm, pos=1] {$\reg{Y}_3$};
      
      \draw[->] ([yshift=-3mm]B1.east) .. controls +(right:20mm) and
      +(left:20mm) .. ([yshift=3mm]B2.west) node [right, pos=0.4]
      {$\reg{Z}_1$};
      
      \draw[->] ([yshift=-3mm]B2.east) .. controls +(right:20mm) and 
      +(left:20mm) .. ([yshift=3mm]B3.west) node [right, pos=0.4]
      {$\reg{Z}_2$};
      
      \draw[->] (leftZ1.east) -- (rightZ1.west)
      node [above, xshift=-5mm, pos=1] {$\reg{Y}_1$}
      node [above, xshift=5mm, pos=0] {$\reg{Y}_1$};

      \draw[->] (leftZ2.east) -- (rightZ2.west)
      node [above, xshift=-5mm, pos=1] {$\reg{Y}_2$}
      node [above, xshift=5mm, pos=0] {$\reg{Y}_2$};

      \draw[->] (leftZ3.east) -- (rightZ3.west)
      node [above, xshift=-5mm, pos=1] {$\reg{Y}_3$}
      node [above, xshift=5mm, pos=0] {$\reg{Y}_3$};

    \end{tikzpicture}
  \end{center}
  \caption{The channel $\Lambda_3^{\ast}=
    (\I_{\Lin(\Y_1\otimes\Y_2)}\otimes\Psi_3^{\ast})
    (\I_{\Lin(\Y_1)}\otimes\Psi_2^{\ast}\otimes\I_{\Lin(\X_3)})
    (\Psi_1^{\ast}\otimes\I_{\Lin(\X_2\otimes\X_3)})$ is
    applied to registers $(\reg{X}_1,\reg{X}_2,\reg{X}_3)$ of a state
    $\rho\in \Density(\Y_1\otimes\Y_2\otimes\Y_3
    \otimes\X_1\otimes\X_2\otimes\X_3)$ with the aim of maximizing the
    fidelity of the output state with the canonical maximally entangled state.}
  \label{fig:online-entanglement-manipulation-1}
\end{figure}
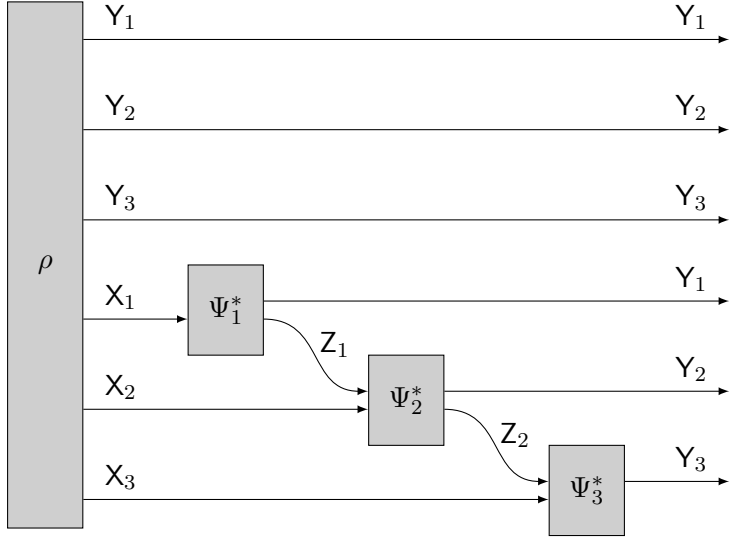
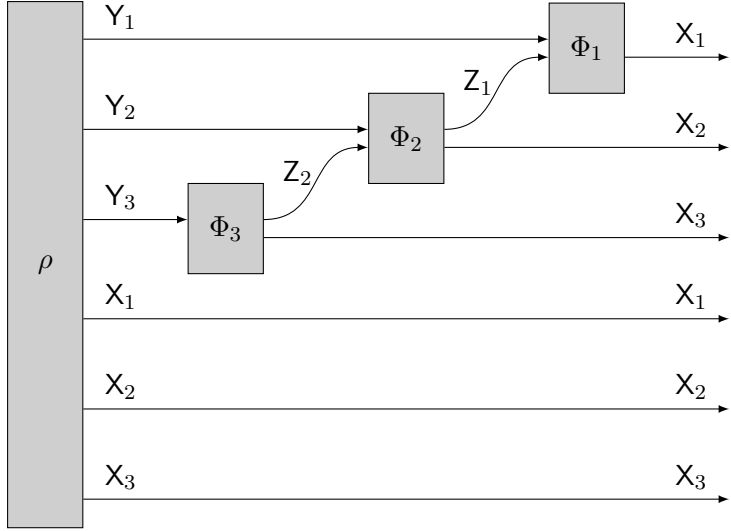
\begin{figure}[p]
  \begin{center}
    \begin{tikzpicture}[scale=0.4, 
        turn/.style={draw, minimum height=12mm, minimum width=10mm,
          fill = ChannelColor, text=ChannelTextColor},
        state/.style={draw, minimum width=10mm, minimum height=70mm,
          fill = ChannelColor},
        invisible/.style={minimum height=1mm, minimum width=10mm},
        >=latex]

      \node (rho) at (-12,-4.2) [state] {$\rho$};

      \node (Y1) at (-12,3) [invisible] {};
      \node (Y2) at (-12,0) [invisible] {};
      \node (Y3) at (-12,-3) [invisible] {};
     
      \node (X1) at (12,3) [invisible] {};
      \node (X2) at (12,0) [invisible] {};
      \node (X3) at (12,-3) [invisible] {};
      
      \node (leftZ1) at (-12,-6) [invisible] {};
      \node (leftZ2) at (-12,-9) [invisible] {};
      \node (leftZ3) at (-12,-12) [invisible] {};
      
      \node (rightZ1) at (12,-6) [invisible] {};
      \node (rightZ2) at (12,-9) [invisible] {};
      \node (rightZ3) at (12,-12) [invisible] {};
     
      \node (B3) at (-6,-3) [turn] {$\Phi_3$};
      \node (B2) at (0,-0) [turn] {$\Phi_2$};
      \node (B1) at (6,3) [turn] {$\Phi_1$};

      \draw[->] ([yshift=3mm]Y3.east) -- ([yshift=3mm]B3.west)
      node [above, xshift=5mm, pos=0] {$\reg{Y}_3$};

      \draw[->] ([yshift=3mm]Y2.east) -- ([yshift=3mm]B2.west)
      node [above, xshift=5mm, pos=0] {$\reg{Y}_2$};

      \draw[->] ([yshift=3mm]Y1.east) -- ([yshift=3mm]B1.west)
      node [above, xshift=5mm, pos=0] {$\reg{Y}_1$};

      \draw[->] ([yshift=-3mm]B3.east) -- ([yshift=-3mm]X3.west)
      node [above, xshift=-5mm, pos=1] {$\reg{X}_3$};

      \draw[->] ([yshift=-3mm]B2.east) -- ([yshift=-3mm]X2.west)
      node [above, xshift=-5mm, pos=1] {$\reg{X}_2$};

      \draw[->] ([yshift=-3mm]B1.east) -- ([yshift=-3mm]X1.west)
      node [above, xshift=-5mm, pos=1] {$\reg{X}_1$};
      
      \draw[->] ([yshift=3mm]B3.east) .. controls +(right:20mm) and
      +(left:20mm) .. ([yshift=-3mm]B2.west) node [left, pos=0.6]
      {$\reg{Z}_2$};
      
      \draw[->] ([yshift=3mm]B2.east) .. controls +(right:20mm) and 
      +(left:20mm) .. ([yshift=-3mm]B1.west) node [left, pos=0.6]
      {$\reg{Z}_1$};
      
      \draw[->] (leftZ1.east) -- (rightZ1.west)
      node [above, xshift=-5mm, pos=1] {$\reg{X}_1$}
      node [above, xshift=5mm, pos=0] {$\reg{X}_1$};

      \draw[->] (leftZ2.east) -- (rightZ2.west)
      node [above, xshift=-5mm, pos=1] {$\reg{X}_2$}
      node [above, xshift=5mm, pos=0] {$\reg{X}_2$};

      \draw[->] (leftZ3.east) -- (rightZ3.west)
      node [above, xshift=-5mm, pos=1] {$\reg{X}_3$}
      node [above, xshift=5mm, pos=0] {$\reg{X}_3$};

    \end{tikzpicture}
  \end{center}
  \caption{A similar process to the one illustrated in
    Figure~\ref{fig:online-entanglement-manipulation-1}, but with channels
    applied to $\reg{Y}_3$, $\reg{Y}_2$, $\reg{Y}_1$ rather than
    $\reg{X}_1$, $\reg{X}_2$, $\reg{X}_3$.}
  \label{fig:online-entanglement-manipulation-2}
\end{figure}

By Theorem~\ref{theorem:main}, one finds that when $\rho$ is pure, the same
optimal value is achieved when the ordering of the channels and the registers
on which they act is reversed, as illustrated in
Figure~\ref{fig:online-entanglement-manipulation-2} for the case $n=3$.
That is, when $\rho$ is a pure state, the optimal value of the semidefinite
program \eqref{eq:SDP-entanglement-concentration} represents the value
\begin{equation}
  \bigip{(\Xi_n\otimes\I_{\Lin(\X_1\otimes\cdots\otimes\X_n)})(\rho)}{
    \vec(\I_{\X_1\otimes\cdots\otimes\X_n})
    \vec(\I_{\X_1\otimes\cdots\otimes\X_n})^{\ast}},
\end{equation}
maximized over all channels $\Xi_n\in\Channel(\Y_1\otimes\cdots\otimes\Y_n,
\X_1\otimes\cdots\otimes\X_n)$ of the form
\begin{equation}
  \Xi_n = \bigl(\Phi_1\otimes\I_{\Lin(\X_2\otimes\cdots\otimes\X_n)}\bigr)
  \cdots
  \bigl(\I_{\Lin(\Y_1\otimes\cdots\otimes\Y_{n-1})}\otimes \Phi_n\bigr)
\end{equation}
for channels $\Phi_1,\ldots,\Phi_n$ taking the form
\begin{equation}
  \Phi_k \in \Channel(\Y_k\otimes\Z_k,\Z_{k-1}\otimes\X_k)
\end{equation}
and for $\Z_2,\ldots,\Z_{n-1}$ arbitrary complex Euclidean spaces
(along with $\Z_0 = \complex$ and $\Z_n = \complex$).
 
\section{Conclusion}

We have identified a time-reversal property for rank-one quantum strategy
functions, explained its connection to conditional min- and max-entropy, and
described an alternative view of this property through an online variant of
pure state entanglement manipulation.
An obvious question arises: are there interesting applications or implications
of this property beyond those we have mentioned?

\subsection*{Acknowledgments}

Yuan Su was supported in part by the Army Research Office (MURI award
W911NF-16-1-0349) and the National Science Foundation (grant 1526380).
John Watrous acknowledges the support of Canada's NSERC.
We thank Fr\'{e}d\'{e}ric Dupuis, James R. Garrison, Brian Swingle, Penghui
Yao, Ronald de Wolf, and M\={a}ris Ozols for helpful discussions, and we thank
the anonymous referees for their comments and suggestions.

\bibliographystyle{plainnat}


\begin{thebibliography}{16}
\providecommand{\natexlab}[1]{#1}
\providecommand{\url}[1]{\texttt{#1}}
\expandafter\ifx\csname urlstyle\endcsname\relax
  \providecommand{\doi}[1]{doi: #1}\else
  \providecommand{\doi}{doi: \begingroup \urlstyle{rm}\Url}\fi

\bibitem[Chiribella and Ebler(2016)]{ChiribellaE16}
G.~Chiribella and D.~Ebler.
\newblock Optimal quantum networks and one-shot entropies.
\newblock \emph{New Journal of Physics}, 18:\penalty0 093053, 2016.
\newblock \doi{10.1088/1367-2630/18/9/093053}.

\bibitem[Chiribella et~al.(2008{\natexlab{a}})Chiribella, D'Ariano, and
  Perinotti]{ChiribellaDP08}
G.~Chiribella, G.~D'Ariano, and P.~Perinotti.
\newblock Quantum circuit architecture.
\newblock \emph{Physical Review Letters}, 101\penalty0 (6):\penalty0 060401,
  2008{\natexlab{a}}.
\newblock \doi{10.1103/PhysRevLett.101.060401}.

\bibitem[Chiribella et~al.(2008{\natexlab{b}})Chiribella, D'Ariano, and
  Perinotti]{ChiribellaDP08b}
G.~Chiribella, G.~D'Ariano, and P.~Perinotti.
\newblock Transforming quantum operations: quantum supermaps.
\newblock \emph{Europhysics Letters}, 83\penalty0 (3):\penalty0 30004,
  2008{\natexlab{b}}.
\newblock \doi{10.1209/0295-5075/83/30004}.

\bibitem[Chiribella et~al.(2009)Chiribella, D'Ariano, and
  Perinotti]{ChiribellaDP09}
G.~Chiribella, G.~D'Ariano, and P.~Perinotti.
\newblock Theoretical framework for quantum networks.
\newblock \emph{Physical Review A}, 80\penalty0 (2):\penalty0 022339, 2009.
\newblock \doi{10.1103/PhysRevA.80.022339}.

\bibitem[Chiribella et~al.(2013)Chiribella, D'Ariano, Perinotti, Schlingemann,
  and Werner]{ChiribellaDPSW13}
G.~Chiribella, G.~D'Ariano, P.~Perinotti, D.~Schlingemann, and R.~Werner.
\newblock A short impossibility proof of quantum bit commitment.
\newblock \emph{Physics Letters A}, 377\penalty0 (15), 2013.
\newblock \doi{10.1016/j.physleta.2013.02.045}.

\bibitem[Datta(2009)]{Datta09}
N.~Datta.
\newblock Min- and max-relative entropies and a new entanglement monotone.
\newblock \emph{IEEE Transactions on Information Theory}, 55\penalty0
  (6):\penalty0 2816--2826, 2009.
\newblock \doi{10.1109/TIT.2009.2018325}.

\bibitem[Eggeling et~al.(2002)Eggeling, Schlingemann, and Werner]{EggelingSW02}
T.~Eggeling, D.~Schlingemann, and R.~Werner.
\newblock Semicausal operations are semilocalizable.
\newblock \emph{Europhysics Letters}, 57\penalty0 (6):\penalty0 782--788, 2002.
\newblock \doi{10.1209/epl/i2002-00579-4}.

\bibitem[Gutoski(2009)]{Gutoski09}
G.~Gutoski.
\newblock \emph{Quantum Strategies and Local Operations}.
\newblock PhD thesis, University of Waterloo, 2009.
\newblock \href{http://hdl.handle.net/10012/4903}{URI: 10012/4903}.

\bibitem[Gutoski and Watrous(2007)]{GutoskiW07}
G.~Gutoski and J.~Watrous.
\newblock Toward a general theory of quantum games.
\newblock In \emph{Proceedings of the 39th Annual ACM Symposium on Theory of
  Computing}, pages 565--574, 2007.
\newblock \doi{10.1145/1250790.1250873}.

\bibitem[Hardy(2012)]{Hardy12}
L.~Hardy.
\newblock The operator tensor formulation of quantum theory.
\newblock \emph{Philosophical Transactions of the Royal Society A},
  370\penalty0 (1971):\penalty0 3385--3417, 2012.
\newblock \doi{10.1098/rsta.2011.0326}.

\bibitem[Kitaev et~al.(2002)Kitaev, Shen, and Vyalyi]{KitaevSV02}
A.~Kitaev, A.~Shen, and M.~Vyalyi.
\newblock \emph{Classical and Quantum Computation}, volume~47 of \emph{Graduate
  Studies in Mathematics}.
\newblock American Mathematical Society, 2002.
\newblock \doi{10.1090/gsm/047/08}.

\bibitem[K\"onig et~al.(2009)K\"onig, Renner, and Schaffner]{KoenigRS09}
R.~K\"onig, R.~Renner, and C.~Schaffner.
\newblock The operational meaning of min- and max-entropy.
\newblock \emph{IEEE Transactions on Information Theory}, 55\penalty0
  (9):\penalty0 4337--4347, 2009.
\newblock \doi{10.1109/TIT.2009.2025545}.

\bibitem[Nielsen and Chuang(2000)]{NielsenC00}
M.~Nielsen and I.~Chuang.
\newblock \emph{Quantum Computation and Quantum Information}.
\newblock Cambridge University Press, 2000.
\newblock \doi{10.1017/CBO9780511976667}.

\bibitem[Watrous(2018)]{Watrous18}
J.~Watrous.
\newblock \emph{The Theory of Quantum Information}.
\newblock Cambridge University Press, 2018.
\newblock \doi{10.1017/9781316848142}.

\bibitem[Wilde(2013)]{Wilde13}
M.~Wilde.
\newblock \emph{Quantum Information Theory}.
\newblock Cambridge University Press, 2013.
\newblock \doi{10.1017/CBO9781139525343}.

\bibitem[Wolkowicz et~al.(2000)Wolkowicz, Saigal, and
  Vandenberge]{WolkowiczSV00}
H.~Wolkowicz, R.~Saigal, and L.~Vandenberge, editors.
\newblock \emph{Handbook of Semidefinite Programming: Theory, Algorithms, and
  Applications}.
\newblock Kluwer Academic Publishers, 2000.
\newblock \doi{10.1007/978-1-4615-4381-7}.

\end{thebibliography}

\end{document}